# Electron spin polarization induced by spin Hall effect in semiconductors with a linear in the momentum spin-orbit splitting of conduction band.

V.L. Korenev

A. F. Ioffe Physical Technical Institute, St. Petersburg, 194021 Russia

*It is shown that spin Hall effect creates uniform spin polarization of electrons in semiconductor with a linear in the momentum spin splitting of conduction band. In turn, the profile of the non-uniform spin polarization accumulated at the edge of the sample oscillates in space even in the absence of an external magnetic field.*

*Corresponding author: korenev@orient.ioffe.rssi.ru

Generation of spin and spin current in non-magnetic semiconductors is based on the spin-orbit coupling of spin with orbital degrees of freedom. Such a coupling leads to the spin Hall effect (SHE) [1] – appearance of the spin current accompanied charge current **j** of unpolarized electrons. In this case electrons with opposite spins deviate in opposite directions perpendicular to **j**. On the one hand, SHE induces an additional current of spin-polarized electrons with spin density **S** perpendicular to both **S** and **j** (anomalous Hall effect). The anomalous Hall effect was found in semiconductors a long time ago [2, 3]. On the other hand, SHE accumulates non-equilibrium spin polarization near the edge (but not in the interior) of the sample. This phenomenon was observed recently [4, 5].

Semiconductors with reduced symmetry (for example, semiconductor quantum wells or strained GaAs) possess a linear in the momentum spin splitting of conduction band that brings about a number of new phenomena. In particular, Dyakonov and Kachorovski (DK) shown [6], that the spin relaxation mechanism of Dyakonov-Perel (DP) [7] is enhanced and become anisotropic in quantum wells. It was predicted and found experimentally [8], that electric current in the quantum well plane is accompanied by the appearance of the effective magnetic field $B_{eff}$ inducing the precession of a mean electron spin **S**. Recently this effect has been reproduced [9] in strained bulk GaAs-type crystals having a linear in the momentum spin splitting of conduction band. As a next step, it was predicted [10] that not only drift but also diffusion charge flow induces average spin-orbit field $B_{eff}$. It has been found recently in strained GaAs [11]. Kalevich-Korenev-Merkulov (KKM) proposed [12] theory on the relationship between the nonequilibrium spin and spin current (spin flux) in a weak spin-orbit coupling regime in crystals with a linear in the momentum spin splitting. They shown that the spin current induces the nonequlibrium spin polarization of electrons all over the crystal. In turn, the spin polarization results in the appearance of spin current. KKM theory presented a unified vision of the DK spin relaxation and precession in averaged effective magnetic fields due to drift-diffusion motion of spin. However it did not take into account the spin Hall effect.

It has been recognized long ago [13] that electric current induces electron spin polarization in systems with a linear in the momentum spin-orbit splitting of conduction band. Spin relaxation process is very essential for this effect [14] that has been interpreted [8] as an "equilibrium" polarization $s_T \sim \mu_B B_{eff}/T$ in magnetic field $B_{eff}$. Current-induced spin polarization was observed recently in strained epilayers of bulk crystals [15] and two-dimensional hole gas [16]. It was stated [15], however, that the observed polarization does not correspond to the theoretical prediction [13, 14] and most likely is the result of the current-induced *generation* of spin rather than spin relaxation process.

Here the KKM theory is generalized to the spin Hall effect. It is shown that SHE generates *uniform* nonequilibrium spin polarization in crystals with a linear in the momentum spin-orbit splitting of conduction band. In turn, the *nonuniform* spin density $S(r)$ near the edge of the sample oscillates in space even in the absence of external magnetic field.

The Hamiltonian of conduction band electron with momentum $\vec{p}$, effective mass m is $\hat{H} = \frac{p^2}{2m} + \hat{s}_\alpha Q_{\alpha\beta} p_\beta$, where the spin-orbit interaction is characterized by a second rank pseudotensor $Q_{\alpha\beta}$ ($\hat{s}_\alpha$ is the operator of the α-component of electron spin) [17]. The spin-orbit interaction can be considered as an interaction of electron spin with effective magnetic field $\vec{B}_p = \hat{Q}\vec{p}/\mu_B g$ ($\mu_B>0$ - Bohr magneton, g – electron g-factor) whose value and direction are determined by those of electron momentum $\vec{p}$. For instance, the spin-orbit interaction of the form $q\hat{\vec{s}}[\vec{p} \times \vec{n}]$ (asymmetrical quantum wells, strained bulk GaAs, wurtzite-type crystals, etc) implies that $Q_{\alpha\beta}=q\varepsilon_{\alpha\beta z}$, with q and $\varepsilon_{\alpha\beta z}$ being spin-orbit constant and Levy-Civita symbol (vector $\vec{n}$ is parallel to z axis) respectively. Steady state spin density $S(r)$ and spin current $J_{\alpha\beta}$ (the $J_{\alpha\beta}$ component of spin current gives the velocity of β –component of spin density along axis α with

α,β=x,y,z) in nondegenerated semiconductors in diffusive regime up to the linear in spin-orbit coupling terms are determined by the unified DP [1] and KKM [12] equations

$$J_{\alpha\beta} = -bE_\alpha S_\beta - D\frac{\partial S_\beta}{\partial x_\alpha} + \beta n \varepsilon_{\alpha\beta\gamma} E_\gamma + \frac{mD}{\hbar}\varepsilon_{\beta j\gamma} Q_{j\alpha} S_\gamma \qquad (1)$$

$$\frac{\partial S_\beta}{\partial t} = -\frac{\partial J_{\alpha\beta}}{\partial x_\alpha} + \frac{m}{\hbar}\varepsilon_{\beta j\gamma} Q_{ji} J_{i\gamma} \qquad (2)$$

The first two terms in Eq. (1) take place without spin-orbit interaction and describe drift of mean spin $\vec{S}$ of electrons (with mobility b) in external electric field $\vec{E}$, and spin diffusion with diffusion coefficient D. The last two terms originate from spin-orbit interaction. The third term gives SHE [1] – spin current induced by the current of charge. It exists even in systems of spherical symmetry and is characterized by parameter β having mobility units. In contrast to it, the forth term in Eq.(1) appears only in crystals with reduced symmetry (for example, nanostructures, strained GaAs-type semiconductors). It describes the spin current emerging in the presence of nonequilibrium electron spin polarization – direct KKM effect [12]. Inverse KKM effect – generation of nonequilibrium spin in the presence of spin current – is given by the second term in Eq. (2). The first term in Eq. (2) describes the change of spin due to inhomogeneous distribution of spin current. Figure 1 illustrates the physics of the direct and inverse KKM effects for the case of asymmetric quantum well with normal $\vec{n}$ along axis **z** and Rashba-type spin-orbit interaction $q\hat{\vec{s}}[\vec{p}\times\vec{n}]$.

Let us demonstrate that SHE generates the uniform nonequilibrium spin density. Consider uniform solution of Eqs. (1,2), i.e. spin density **S** does not vary in space. Substitution of Eq. (1) into Eq. (2) gives [18]

$$\frac{\partial \vec{S}}{\partial t} = -n\frac{m\beta}{\hbar}\hat{Q}^T\vec{E} - \hat{\Gamma}\vec{S} + \vec{\Omega}_{eff}\times\vec{S} \qquad (3)$$

The second term in Eq.(3) gives Dyakonov-Kachorovsky spin relaxation rate [6] with relaxation tensor $\Gamma_{\alpha\beta} = \Gamma_{\beta\alpha} = Dm^2[Sp(\hat{Q}\hat{Q}^T)\delta_{\alpha\beta} - (\hat{Q}\hat{Q}^T)_{\alpha\beta}]/\hbar^2$. The third term describes precession of mean spin in effective magnetic field $\vec{B}_{eff} = \hat{Q}\vec{p}_{dr}/\mu_B g$ with Larmor frequency $\vec{\Omega}_{eff} = \hat{Q}\vec{p}_{dr}/\hbar$ [8] ($\vec{p}_{dr} = -mb\vec{E}$ is the drift momentum of electron ensemble). In contrast to these, the first term of Eq. (3) represents the *uniform generation* of nonequilibrium spin by electric current in the SHE conditions. It should be noted that there is "intrinsic" [2, 19] (that exists even in the absence of scattering) и "extrinsic" [1] (based on a skew scattering) SHE mechanisms. Respectively to it "intrinsic" and "extrinsic" contributions into generation term $\dot{\vec{S}}_g = -nm\beta Q^T\vec{E}/\hbar$ are present. In particular case $Q_{\alpha\beta}=q\varepsilon_{\alpha\beta z}$, the SHE creates electron spin with generation rate $\dot{\vec{S}}_g = -n\vec{\Omega}_{eff}\beta/b$ antiparallel (under β>0) to vector $\vec{\Omega}_{eff} = -mbq(\vec{E}\times\vec{n})/\hbar$ and directed in the QW plane perpendicular to the electric field. Figure 2 illustrates the physics of the SHE-induced generation effect. The steady-state value of spin density $\vec{S} = \dot{\vec{S}}_g\tau_s = -n\beta\vec{\Omega}_{eff}\tau_s/b$ is the larger, the longer DK spin relaxation time $\tau_s = \Gamma_{xx}^{-1} = \Gamma_{yy}^{-1} = \hbar^2/Dm^2q^2$. Thereby this effect is inherently different from spin orientation by electric current due to spin relaxation in effective magnetic field [13, 14]. External magnetic field $\vec{B}$ brings about the precession of the mean spin with Larmor frequency $\vec{\Omega}$. In steady state conditions vector $\vec{S}$ is rotated by the angle $\Omega\tau_s$ (the Hanle effect). To estimate the value of the spin polarization we use the quantity $\beta/b \sim 10^{-3}$ from the experiment [4]. Putting $\Omega_{eff}\tau_s \sim 0.1$ we get $S/n \sim 10^{-4}$. Generation of the uniformly distributed nonequilibrium spin of such a value and its Larmor rotation was observed in the paper [15] pointing to the difference between the origin of the effect and theoretical prediction [13, 14]. In contrast, this model explains naturally the result [15]. SHE and the generation of spin by electric current have *the same* origin.

The mentioned above ability for the nonequilibrium spin to be accumulated gives rise to the interesting consequence. Substituting the steady-state spin density $\vec{S} = -n\beta\vec{\Omega}_{eff}\tau_s/b$ into Eq. (1) for the spin current we obtain that the last two terms compensate each other and SHE disappears. Figuratively speaking, the SHE is converted *entirely* into the electron spin polarization. Such compensation is not related with the specific type of spin-orbit interaction.

Indeed, as it follows from Eq. (2), in steady-state regime the second term in Eq. (2) is equal to zero in case of uniform polarization. This is certainly true, if $J_{\alpha\beta} = 0$. Then the SHE and direct KKM effect (the last two terms in Eq.(1)) compensate each other in a linear in electric field approximation for sufficiently general matrix $\hat{Q}$. As a result the accumulation of nonuniform spin polarization near the edge of the sample is absent. In reality, however, the full compensation of SHE will not take place. One possibility is that the Eq. (2) should take into account spin loss due to the processes not related with linear in momentum terms (such as cubic in momentum terms, hyperfine interaction etc.). The presence of additional relaxation channels implies that $J_{\alpha\beta} \neq 0$ in the bulk of the sample anymore. At the same time the edge spin current component normal to the boundary is zero, i.e. $J_{n\beta} = 0$ (n-th component means the flow perpendicular to the boundary). Alternative possibility includes surface effects into Eq. (1). For example, spin loss at the edge is different from that of the bulk. It implies that the spin current at the boundary $J_{n\beta} \neq 0$, whereas it is still zero in the bulk. In both cases SHE accumulates the nonequilibrium polarization near the edge of the sample as before.

Now, we show that the SHE-induced nonuniform spin polarization at the edge of the sample oscillates in crystals with a linear in momentum spin-orbit term. In the nonuniform case there is additional contribution into Eq. (3)

$$b(\vec{E}\nabla)\vec{S} + D\nabla^2\vec{S} - \frac{2mD}{\hbar}(\hat{Q}\nabla \times \vec{S}) \qquad (3a)$$

The first two terms in Eq. (3a) describe the usual drift-diffusion motion of average spin density. The last term has spin-orbit origin [20]. It is responsible for the oscillations of $\vec{S}$ even in the absence of external magnetic field. Consider, for example, nondegenerated two-dimensional electron gas in asymmetrical quantum well ($Q_{\alpha\beta}=q\varepsilon_{\alpha\beta z}$). Let sample be elongated along **x** axis and electric field $\vec{E}\|x$ (Fig.2). The distribution of spin polarization at the left edge of the sample (y=0) depends on coordinate **y** only. It is determined by Eqs.(1, 3, 3a) and boundary condition $J_{y\beta} = 0$. In linear approximation we can omit the terms quadratic in electric field in Eqs.(1, 3, 3a) because the spin density is generated by $\vec{E}$. In this case the mean spin turns over in the (yz) plane. For the SHE to exist, we include into Eqs.(3, 3a) the leakage of spin different from the DK relaxation as stated above. In primary case such a term takes the form $-\vec{S}/\tau'_s$ with characteristic time $\tau'_s$. The steady state Eqs. (3,3a) are reduced to

$$\frac{d^2 S_y}{dy^2} + \frac{2}{\ell}\frac{dS_z}{dy} - \frac{S_y + S_0}{\ell^2} - \frac{S_y}{L^2} = 0; \quad \frac{d^2 S_z}{dy^2} - \frac{2}{\ell}\frac{dS_y}{dy} - \frac{2S_z}{\ell^2} - \frac{S_z}{L^2} = 0 \qquad (5)$$

with boundary conditions at y=0: $\frac{dS_y}{dy} + \frac{S_z}{\ell} = 0$ и $-\frac{dS_z}{dy} + \frac{S_y + S_0}{\ell} = 0$. Two parameters - $S_0 = n\beta E\ell/D$ and $\ell = \hbar/mq$ - give the value of the induced spin polarization and characteristic length. Parameter $L^2 = D\tau'_s$ results from other spin relaxation mechanisms. To derive the Eq. (5) we used the nonzero components of relaxation tensor $\hat{\Gamma}$: $\Gamma_{yy} = \Gamma_{zz}/2 = D/\ell^2 \equiv 1/\tau_s$. If the DK mechanism dominates ($\tau'_s \gg \tau_s$) then the same parameter $\ell$ determines both precession period in space and decay length. Equations (5) enable analytical solution that we do not present here due to awkwardness. However, behavior of spin density is clear and without formulas: it oscillates (with maximal amplitude $\sim S_0 \tau_s/\tau'_s$) and saturates at a level $S_y \approx -S_0$ ($S_z = 0$) deep into the sample. As an illustration, Fig. 3 shows the out-of-plane $S_z(y)$ component profile for $\tau'_s \gg \tau_s$. The opposite edge of the sample demonstrates inversion of $S_z(y)$ in accord with the overall tendency of SHE [1].

It is shown that the spin Hall effect generates uniform nonequilibrium spin polarization in the bulk of the crystals with a linear in momentum spin splitting of conduction band. In turn, the nonuniform spin density profile near the edges of the sample oscillates in space even in the absence of external magnetic field. The results are valid not only for quantum wells but for bulk crystals whose symmetry allows the linear in momentum spin splitting.

Author is grateful to E.L. Ivchenko and I.A. Merkulov for fruitful discussions, M.V. Lazarev for the help in manuscript preparation. The paper is supported by CRDF, RSSI, RFBR grants, the Department of Physical Sciences and the Presidium of the Russian Academy of Sciences.

Figure Captions

Figure 1. Illustration of the direct (a) and inverse (b) KKM effects [12] for the spin-orbit interaction $q\hat{\vec{s}}(\vec{p}\times\vec{n})$.

(a) Spin polarization $S_z \neq 0$ of electron ensemble induces spin current (here all spins look up, symbol ⊙ ). Electrons with oppositely directed momenta $\vec{p}$ and $-\vec{p}$ acquire oppositely directed spin components $\Delta\vec{S}_p$ and $\Delta\vec{S}_{-p}$ as a result of spin precession in effective magnetic field with frequency $\vec{\Omega}_p = q(\vec{p}\times\vec{n})/\hbar$. Such a correlation between spin and momentum implies the non-zero flow of y-th spin component into **y** direction, i.e. non-zero spin current component $J_{yy} \propto \langle p_y \Delta S_y \rangle \neq 0$ (angle bracket denote ensemble averaging).

(b) Spin current generates spin polarization of electrons. Suppose a spin current $J_{yz}$ of z-th spin component in **y** direction in the ensemble of initially unpolarized electrons: one half of electrons possess spin up and momentum $\vec{p}$, whereas another half – spin down (⊗) and momentum $-\vec{p}$. Then spin precession with $\vec{\Omega}_p$ and $\vec{\Omega}_{-p}$ frequencies generates spin at a rate $\dot{\vec{S}}$ opposite to axis **y**.

Figure 2. Spin Hall effect generates electron spin for the spin-orbit interaction of $q\hat{\vec{s}}(\vec{p}\times\vec{n})$ type. Electric field $\vec{E}\|x$ brings about the drift of electrons opposite to it. Electrons with opposite spins deflect in opposite directions perpendicular to the field due to SHE [1]. Two typical trajectories are shown. Spins of electrons at every trajectory rotate about $\vec{\Omega}_p$ ($\vec{\Omega}_{p'}$) direction leading to the generation of spin with a rate $\dot{\vec{S}}_p$ ($\dot{\vec{S}}_{p'}$). Averaging over trajectories gives the mean effective magnetic field $\vec{B}_{eff}$ [8], bringing about precession with frequency $\vec{\Omega}_{eff} = -mbq(\vec{E}\times\vec{n})/\hbar$, and mean spin generation rate $\dot{\vec{S}}_g$ antiparallel to vector $\vec{\Omega}_{eff}$.

Figure 3. Steady state distribution of z-th component of the electron spin density $S_z(y)$ near the edges of the sample in the spin Hall effect conditions.

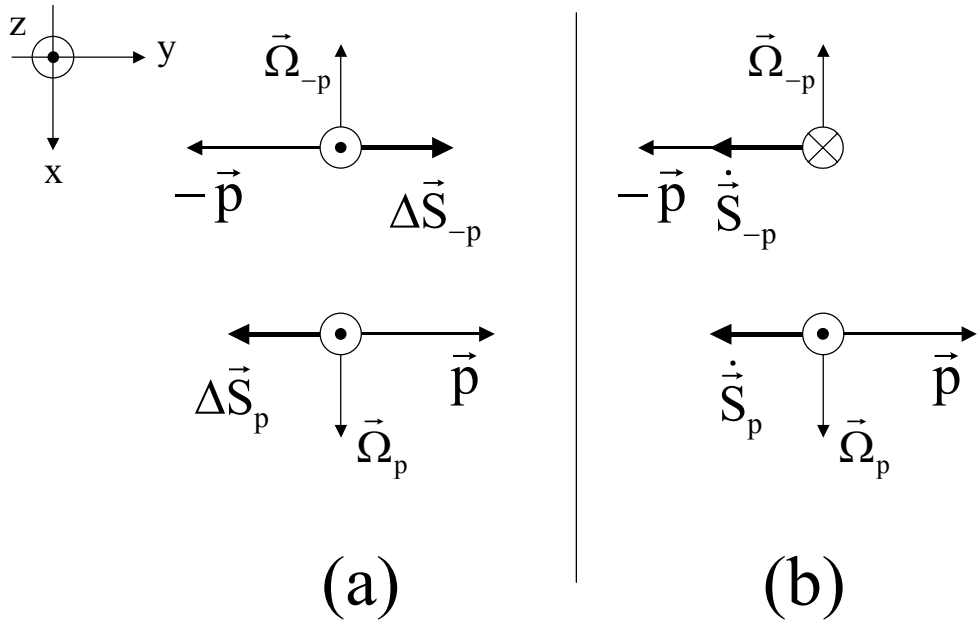

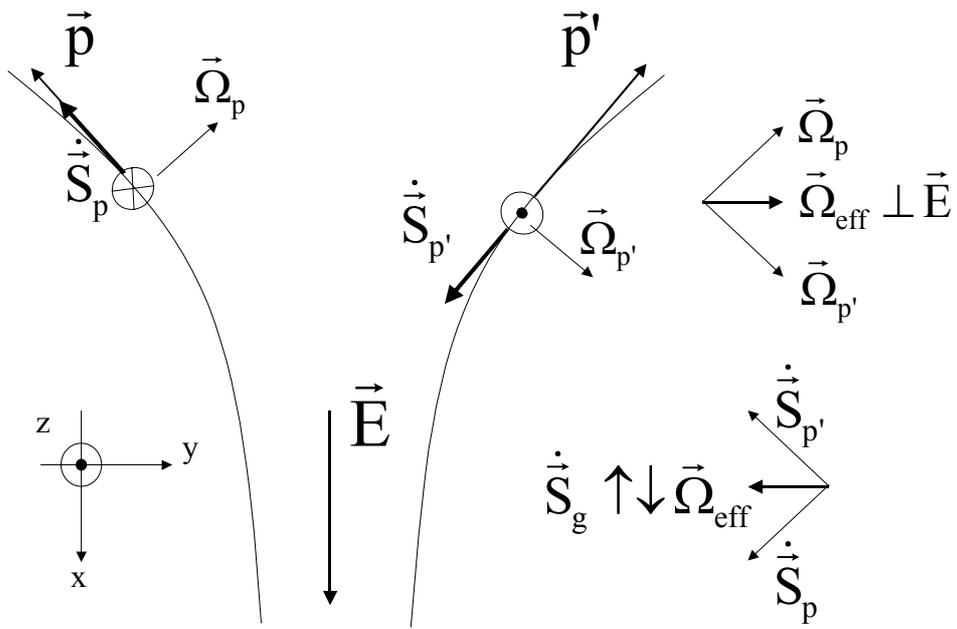

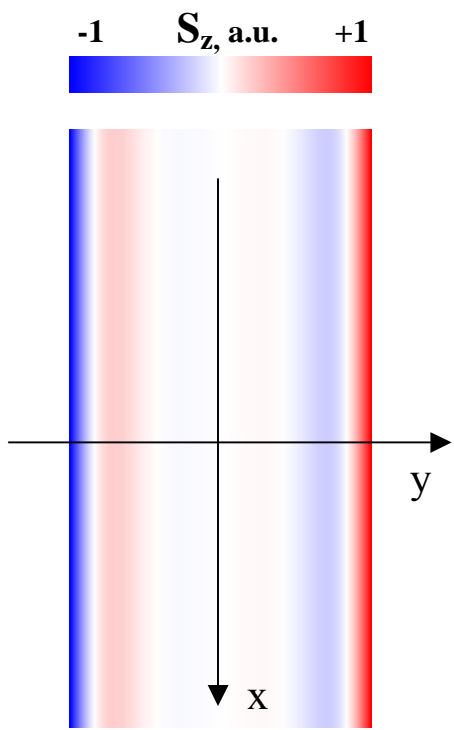